# Spectroscopy of Charge Fluctuators Coupled to a Cooper Pair Box

F. C. Wellstood,[1,2] Z. Kim[2,3] and B. Palmer[3]

[1.] Joint Quantum Institute and Center for Nanophysics and Advanced Materials

[2.] Department of Physics, University of Maryland, College Park, Maryland 20742

[3.] Laboratory for Physical Sciences, College Park, MD 20740

**Abstract**

We analyze the behavior of a Cooper Pair Box (CPB) that is coupled to charge fluctuators that reside in the dielectric barrier layer in the box's ultra-small tunnel junction. We derive the Hamiltonian of the combined system and find the coupling between the CPB and the fluctuators as well as a coupling between the fluctuators that is due to the CPB. We then find the energy levels and transition spectrum numerically for the case of a CPB coupled to a single charge fluctuator, where we treat the fluctuator as a two-level system that tunnels between two sites. The resulting spectra show the usual transition spectra of the CPB plus distinctive transitions due to excitation of the fluctuator; the fluctuator transitions are 2-$e$ periodic and resemble saw-tooth patterns when plotted as a function of the gate voltage applied to the box. The combined CPB-fluctuator spectra show small second-order avoided crossings with a size that depends on the gate voltage. Finally, we discuss how the microscopic parameters of the model, such as the charge times the hopping distance, the tunneling rate between the hopping sites, and the energy difference between the hopping sites, can be extracted from CPB spectra, and why this yields more information than can be found from similar spectra from phase qubits.

Two-level fluctuators (TLF) of one kind or another have long been recognized as an underlying cause of resistance fluctuations in metals [1,2], critical current fluctuations in Josephson junctions [3-7], excess flux noise in SQUIDs [7-9], and telegraph noise and excess charge noise in Coulomb blockade devices [10-13]. Not surprisingly, different types of fluctuators lead to different types of fluctuations. For example, observations of excess flux noise in high-$T_c$ superconductors have revealed that the two-level fluctuators are just vortices that hop between different pinning sites [14-16]. In other cases, the microscopic nature of the fluctuators is still unclear. For example, the unusual behavior of flux noise at temperatures below 1 K appears to be inconsistent with the hopping of vortices and may be due to fluctuating electron spins [17-18]. The situation for charge and critical current fluctuators is somewhere between these extremes. After more than two decades of research, some basic information is known about the microscopic fluctuators responsible for charge and critical current fluctuations - the fluctuators appear to be moving charges or rotating electric dipoles in the insulating junction barrier or nearby dielectric layers [3] - but the precise microscopic identification of the fluctuators is not settled.

There are several reasons why it has been difficult to be certain of the precise microscopic agents causing charge or critical current fluctuations. First, the effects from individual fluctuators are typically very small, especially for the cryogenic or milli-Kelvin temperatures of interest here, and this typically makes the resulting experiments quite challenging. Second, a variety of materials and processes have been used to build devices, and the presence of a fluctuator may well depend on both the materials and the fabrication technique. Third, while measurements of telegraph noise can reveal microscopic information about individual fluctuators, in many cases it is possible only to distinguish the largest one or two fluctuators in a background noise of smaller fluctuations. Compared to the number of atoms in even the smallest device, such observable discrete fluctuators are extremely rare and thus may not be representative of a typical atomic scale defect in the device. Finally, much of the experimental data obtained on fluctuators consists of relatively smooth *1/f* noise power spectra. Such noise spectra arise from a distribution of many fluctuators [1,19] and it is often not possible to resolve individual fluctuators. From smooth spectra, it is difficult to determine uniquely such basic microscopic parameters as the hopping distance or the absolute number of fluctuators in a given energy range.

Research in quantum computing based on superconducting devices has led to increased interest in understanding two-level fluctuators. In qubit research, fluctuators can be a serious problem because they can lead to decoherence, dissipation, inhomogenous broadening and a decrease in measurement fidelity [20-23]. With this new interest have also come new approaches to the problem. In particular, microwave spectroscopy of Josephson phase qubits and flux qubits has revealed the presence of small un-intended avoided crossings in the transition spectrum due to coupling between the device and individual two-level fluctuators [20-23]. Similar avoided crossings have also been observed recently [23] in the transition spectra of Cooper pair boxes (CPB), allowing the new spectroscopic approach to be applied to charge fluctuators in a system that is a sensitive detector of charge.

Here, we examine the behavior of a Cooper pair box [25-29] that is coupled to charged two-level fluctuators and show how microscopic information can be extracted about the fluctuators. We first summarize the basic physics of the Cooper pair box. We then derive the Hamiltonian for a system where charged fluctuators are coupled to a CPB and find the energy levels numerically for the case of a single two-level fluctuator coupled to a CPB. We find that the resulting transition spectra show a distinct saw-tooth feature that is characteristic of the fluctuator. In addition, we find that three key model parameters for individual fluctuators (the charge times the hopping distance, the tunneling rate between the hopping sites, and the energy difference between the hopping sites) can be uniquely extracted from the spectra. Finally, we conclude with a summary of our key results and discuss how the CPB-fluctuator spectra provide additional information about the fluctuators that is not found in analogous spectra from phase qubits, suggesting that measurements of CPB-fluctuator spectra may aid in the identification of microscopic charge fluctuators.

**Charging Energy of the Cooper Pair Box**

In this section we summarize the energy considerations for a Cooper pair box when no charge fluctuators are present. Considering the circuit schematic of the Cooper pair box shown in Fig. 1(a), the charging energy of the capacitors in the CPB without any charge fluctuators is [30]:

$$U = \frac{1}{2}C_j(V_i - 0)^2 + \frac{1}{2}C_g(V_g - V_i)^2 \qquad (1)$$

where $C_j$ is the capacitance of the ultra-small junction, $C_g$ is the capacitance of the gate electrode, $V_i$ is the island potential, and $V_g$ is the gate voltage. The island potential $V_i$ is determined by $V_g$ and the number $n$ of excess Cooper pairs on the island. Each Cooper pair has charge $2e$; note that here $e = -1.609 \times 10^{-19}$ C is negative, as in ref. [30]. One finds for the island potential [30]:

$$V_i = \frac{C_g}{C_j + C_g} V_g + \frac{2ne}{(C_j + C_g)} = \frac{e}{C_\Sigma}(2n - n_g) \qquad (2)$$

where:

$$C_\Sigma = C_j + C_g \qquad (3)$$

is the total capacitance of the island and $n_g = -C_g V_g/e$ is the gate number. Also, in writing Eq. (2)

we have implicitly assumed that there are no quasiparticles present.

To properly describe the systems behavior, we need to take into account the work $W$ done by the gate voltage source when the number $n$ of Cooper pairs on the island changes. The appropriate quantity is the charging energy minus the work:

$$E = U - W. \tag{4}$$

To find the work, we note that if $n$ excess pairs have accumulated on the island, a charge of

$$Q_g = -2ne\frac{C_g}{C_\Sigma} \tag{5}$$

will be attracted onto the gate electrode (the plate of the gate capacitor $C_g$ that is directly attached to the gate voltage source $V_g$). Since this charge has to be supplied by the gate voltage source at potential $V_g$, the total work the source has to do is $Q_g V_g$ and thus:

$$W = -2ne\frac{C_g}{C_\Sigma}V_g = 2nn_g\frac{e^2}{C_\Sigma}. \tag{6}$$

After some algebraic manipulation, one finds that the charging energy minus the work is:

$$E = E_c(2n - n_g)^2 + E_c\left(\frac{C_j}{C_g} - 1\right)n_g^2 \tag{7}$$

where $E_c = e^2/2C_\Sigma$. Since the second term in Eq. (7) does not depend on $n$, it will not affect the dynamics and can be dropped [30].

**Analysis of a CPB when Charge Fluctuators are Present**

We now consider the case where two charge fluctuators are coupled to the Cooper pair box. The result for two charges can easily be generalized to more charges and is sufficient to reveal that the CPB will generate a coupling term between any two fluctuators. For this case, we again need to find the charging energy minus the work done by the gate voltage source. We can

write:

$$E = \frac{1}{2}C_j V_i^2 + \frac{1}{2}C_g(V_g - V_i)^2 - W - \sum_{j=1}^{2} W_j \qquad (8)$$

where $W_j$ is the work done by the gate voltage source when the $j$-th fluctuator moves. Here, $W$ is the work done by the gate voltage source when n changes and is given by Eq. (6).

To proceed, we assume that the two charge fluctuators are in the ultra-small tunnel junction barrier, *i.e.* in the capacitor $C_j$, where the device is most sensitive to charge motion [31-32]. The charges, of strength $Q_1$ and $Q_2$ respectively, will induce charges $Q_{p1}$ and $Q_{p2}$ on the island given by [31-32]:

$$Q_{pj} = -\frac{x_j}{d}Q_j \qquad (9)$$

where $x_j$ is the distance of the $j$-th charge from the ground plate of the ultra-small junction and $d$ is the thickness of the junction's oxide barrier [see Fig. 1(b)]. $Q_{p1}$ and $Q_{p2}$ are polarization charges that are localized on the island close to the charges $Q_1$ and $Q_2$. Since the net charge on the island will be constant for n constant, a compensating charge of $-(Q_{p1}+Q_{p2})$ must exist on the rest of the island, distributed proportionally over the gate capacitor and junction capacitor. This in turn induces a charge $Q_g$ on the gate electrode given by:

$$Q_g = -\left(\frac{x_1}{d}Q_1 + \frac{x_2}{d}Q_2\right)\frac{C_g}{C_\Sigma} \qquad (10)$$

This charge must be delivered by the gate voltage source, and requires work:

$$\sum_{j=1}^{2} W_j = Q_g V_g = \frac{e}{C_\Sigma}\left(\frac{x_1}{d}Q_1 + \frac{x_2}{d}Q_2\right)n_g \qquad (11)$$

From similar considerations, we find the island potential $V_i$ is given by:

$$V_i = \frac{1}{C_\Sigma}[C_g V_g + 2ne] + \frac{1}{C_\Sigma}\left(\frac{x_1}{d}Q_1 + \frac{x_2}{d}Q_2\right) \qquad (12)$$

From Eqs. (6), (8), (11) and (12), we find:

$$E = E_c \left(2n - n_g + \sum_{j=1}^{2} \frac{Q_j}{e} \frac{x_j}{d}\right)^2 + E_c \left(\frac{C_j}{C_g} - 1\right) n_g^2. \tag{13}$$

Comparing Eqs. (7) and (13), we see that the effect of charge motion is similar to a shift in the gate voltage and, again, the second term in Eq. (13) has no effect on the dynamics and can be dropped.

We can now construct the Hamiltonian H for the combined system by taking Eq. (13) and adding the kinetic energy of the charge fluctuator, the Josephson coupling energy of the ultra-small tunnel junction, and the local potential experienced by the charge fluctuators. We can write:

$$H = H_{CPB} + \sum_{j=1}^{2} H_j + \sum_{j=1}^{2} H_{CPB-j} + H_{12} \tag{14}$$

The first term in Eq. (14):

$$H_{CPB} = E_c (2n - n_g)^2 - E_j \cos(\gamma) \tag{15}$$

is the Hamiltonian of the Cooper pair box [26] with no charge fluctuators, where $E_j$ is the Josephson energy and $\gamma$ is the gauge invariant phase difference across the junction [31]. The second term in Eq. (14) is a sum over $H_j$, where

$$H_j = \frac{\bar{p}_j^2}{2m_j} + \frac{Q_j^2}{2C_\Sigma} \left(\frac{x_j}{d}\right)^2 + U_j(\bar{r}_j) \tag{16}$$

is the Hamiltonian of the $j$-th charge, which has charge $Q_j$, position $\bar{r}_j$, momentum $\bar{p}_j$, mass $m_j$, and moves in a local potential $U_j(\bar{r}_j)$. The third term in Eq. (14):

$$H_{CPB-j} = 2E_c (2n - n_g) \frac{Q_j}{e} \left(\frac{x_j}{d}\right) \tag{17}$$

is the coupling between the CPB and the *j*-th charge. Finally,

$$H_{12} = U_{12}(x_1, x_2) + \frac{Q_1 Q_2}{C_\Sigma} \frac{x_1 x_2}{d^2} \tag{18}$$

describes the interaction between the two charges. The potential $U_{12}$ accounts for any direct local interactions between the two charges. If the charge defects are few in number and randomly distributed in the dielectric, then we expect $U_{12}$ will typically be small because the junction electrodes will screen electric fields on a length scale given by $d/2\pi \sim 0.2$ nm. On the other hand, the term $Q_1 Q_2 x_1 x_2 / C_\Sigma d^2$ in Eq. (18) came from the charging energy, as can be seen by expanding the first term in Eq. (13). This is an indirect electrostatic interaction between two fluctuators due to the capacitor plates and does not directly depend on the separation between the charges.

**Two-state Model of a Charge Fluctuator**

To illustrate the behavior of a CPB coupled to a charge fluctuator, we will further simplify the situation and consider a CPB that is coupled to just one fluctuator that can tunnel between two locations. In this section, we first summarize the properties of such a two-level fluctuator (TLF) when the coupling to the CPB is ignored. From Eq. (16), we can write the Hamiltonian of the fluctuator as:

$$H_1 = \frac{\bar{p}_1^2}{2m_1} + U_1(\bar{r}_1) \tag{19}$$

where we have absorbed the quadratic potential term $Q_1^2 x_1^2 / 2C_\Sigma d^2$ that occurs in Eq. (16) into the arbitrary potential $U_1$.

For the charge fluctuator to behave as a two-level fluctuator, we will assume that the potential $U_1(\bar{r}_1)$ has two wells that the fluctuator can tunnel between [13,32], as shown in Fig. 1(c). In the absence of tunneling, we will assume the energy of the fluctuator is $E_a$ in well *a* and $E_b$ in well *b*, and that the corresponding distances from the ground electrode are $x_a$ and $x_b$. With these simplifications, $H_1$ can be written as a 2x2 matrix:

$$H_1 = \begin{pmatrix} E_a & T_{ab} \\ T_{ab} & E_b \end{pmatrix} \quad (20)$$

where $T_{ab}$ determines the tunneling rate between the two wells. Here we have chosen basis states:

$$|a> = \begin{pmatrix} 1 \\ 0 \end{pmatrix} \qquad |b> = \begin{pmatrix} 0 \\ 1 \end{pmatrix} \quad (21)$$

where |a> corresponds to the fluctuator being localized in well *a* and |b> to the fluctuator being localized in well *b*.

The two eigenstates of the isolated fluctuator can be found from Eq. (20). To simplify the resulting expressions, we set $E_a = 0$. We can then write the ground state as:

$$|g_f> = \frac{1}{\sqrt{\left(\frac{1}{2}\left(E_b - \sqrt{E_b^2 + 4T_{ab}^2}\right)\right)^2 + T_{ab}^2}} \left\{ T_{ab}|a> + \frac{1}{2}\left(E_b - \sqrt{E_b^2 + 4T_{ab}^2}\right)|b> \right\} \quad (22)$$

and the excited state as:

$$|e_f> = \frac{1}{\sqrt{\left(\frac{1}{2}\left(E_b - \sqrt{E_b^2 + 4T_{ab}^2}\right)\right)^2 + T_{ab}^2}} \left\{ -\frac{1}{2}\left(E_b - \sqrt{E_b^2 + 4T_{ab}^2}\right)|a> + T_{ab}|b> \right\} \quad (23)$$

with corresponding energies:

$$E_{gf} = \frac{1}{2}\left(E_b - \sqrt{E_b^2 + 4T_{ab}^2}\right) \quad (24)$$

$$E_{ef} = \frac{1}{2}\left(E_b + \sqrt{E_b^2 + 4T_{ab}^2}\right). \tag{25}$$

From Eqs. (20)-(25) we can determine all of the properties of the isolated fluctuator. For example, the mean position of the charge when it is in one or the other of the eigenstates is:

$$<g_f|x_j|g_f> = \frac{1}{\left(\frac{1}{2}\left(E_b - \sqrt{E_b^2 + 4T_{ab}^2}\right)\right)^2 + T_{ab}^2}\left\{T_{ab}^2 x_a + \frac{1}{4}\left(E_b - \sqrt{E_b^2 + 4T_{ab}^2}\right)^2 x_b\right\} \tag{26}$$

$$<e_f|x_1|e_f> = \frac{1}{\left(\frac{1}{2}\left(E_b - \sqrt{E_b^2 + 4T_{ab}^2}\right)\right)^2 + T_{ab}^2}\left\{\frac{1}{4}\left(E_b - \sqrt{E_b^2 + 4T_{ab}^2}\right)^2 x_a + T_{ab}^2 x_b\right\} \tag{27}$$

The corresponding charge shift induced on the island when the fluctuator makes a transition from the ground state to the excited state is:

$$\Delta Q_p = -\frac{<e_f|x|e_f>}{d}Q + \frac{<g_f|x|g_f>}{d}Q \tag{28}$$

Substituting for the states we find:

$$\Delta Q_p = \left(\frac{Q(x_b - x_a)}{d}\right)\frac{\left(E_b - \sqrt{E_b^2 + 4T_{ab}^2}\right)^2 - 4T_{ab}^2}{\left(E_b - \sqrt{E_b^2 + 4T_{ab}^2}\right)^2 + 4T_{ab}^2}. \tag{29}$$

We note that for $T_{ab} >> E_b$, Eq. (29) gives $\Delta Q_p = 0$. This is what one expects. When the tunneling is very large, the fluctuator will have nearly the same amplitude to be in well *a* as in well *b*. This will be true for both the ground state and the excited state, and implies that there will little change in the average position of the ion or corresponding change in the induced charge on the island when the state changes. On the other hand, in the limit $T_{ab} << E_b$, one finds $\Delta Q_{pg} = Q\Delta x/d$, which is the full amount expected from a charge $Q$ shifting its position by $\Delta x = x_b - x_a$. This difference in behavior for large and small tunneling implies an interesting *disconnection* between charge noise and charge-induced microstate splittings. Charge noise requires a change in the induced charge and so will be produced most effectively by fluctuators with relatively smaller $T_{ab}/E_b$. Of course a small $T_{ab}/E_b$ would not necessarily prevent fluctuations because movement between the wells could still be driven via thermal activation over the fluctuator barrier. On the other hand, as discussed in the next section, large $T_{ab}/E_b$ tends to produce large splittings in the CPB spectra.

**Energy levels and transition spectrum of a CPB coupled to a charged TLF**

We now consider a system that has a CPB and just one charge fluctuator in the tunnel barrier of the CPB. Equation (14) then reduces to:

$$H = \left\{ E_c (2n - n_g)^2 - E_j \cos(\gamma) \right\} + \left\{ \frac{\bar{p}_1^2}{2m_1} + \frac{Q_1^2}{2C_\Sigma} \left( \frac{x_1}{d} \right)^2 + U_1(\bar{r}_1) \right\} + \frac{e^2}{C_\Sigma} (2n - n_g) \frac{Q_1}{e} \left( \frac{x_1}{d} \right) \quad (30)$$

The first term in brackets is just $H_{CPB}$ and the second term in brackets is $H_1$, the Hamiltonian of a charge fluctuator when it is not coupled to the CPB. As described in the previous section we will replace this term with $H_1$ given by Eq. (20) to produce a two-level fluctuator.

We next choose the basis states of the CPB to be the charge eigenstates of the box. Since the resulting expressions are quite cumbersome if more than two levels are used, for simplicity we will restrict ourselves to the subspace spanned by the charge states corresponding to $n=0$ and 1 excess Cooper pairs on the island. These states are a reasonable choice for a basis for finding the ground state $|g_b\rangle$ of the box if $E_c >> E_j$ and the gate number is in the range of about $0 < n_g < 2$. On the other hand, with the restriction to $n=0$ and 1, we will only achieve a good representation of the first excited state $|e_b\rangle$ in the range of about $1/2 < n_g < 3/2$. We would need to use $n = -1, 0, 1$ to cover the range $0 < n_g < 1$, and $n = 0, 1, 2$ to cover the range $1 < n_g < 2$. Nevertheless, with this two state picture of the CPB, we can then write the uncoupled CPB Hamiltonian in matrix form as [26]:

$$H_{CPB} = \begin{bmatrix} E_c(0-n_g)^2 & -\dfrac{E_j}{2} \\ -\dfrac{E_j}{2} & E_c(2-n_g)^2 \end{bmatrix} \qquad (31)$$

We now choose the basis states for the combined system (CPB and TLF) as the charge-position states $|n,m\rangle$, where $n = 0, 1$ is the charge state of the box and $m$ is $a$ or $b$, corresponding to the position of the fluctuator. We can write the basis states of the combined system explicitly as:

$$|0,a\rangle = \begin{pmatrix}1\\0\\0\\0\end{pmatrix} \quad |1,a\rangle = \begin{pmatrix}0\\1\\0\\0\end{pmatrix} \quad |0,b\rangle = \begin{pmatrix}0\\0\\1\\0\end{pmatrix} \quad |1,b\rangle = \begin{pmatrix}0\\0\\0\\1\end{pmatrix} \qquad (32)$$

For convenience, we will take $x_a=0$ and $E_a=0$. With this choice of basis, the Hamiltonian for the combined system of the CPB and charge fluctuator can be written explicitly in matrix form as:

$$H = \begin{pmatrix} E_c(0-n_g)^2 & -\dfrac{E_j}{2} & T_{ab} & 0 \\ -\dfrac{E_j}{2} & E_c(2-n_g)^2 & 0 & T_{ab} \\ T_{ab} & 0 & E_c(0-n_g)^2 + E_b + E_r(0-n_g) & -\dfrac{E_j}{2} \\ 0 & T_{ab} & -\dfrac{E_j}{2} & E_c(2-n_g)^2 + E_b + E_r(2-n_g) \end{pmatrix} \qquad (33)$$

where:

$$E_r = 2\frac{Q}{e}\frac{\Delta x}{d}E_c \qquad (34)$$

sets the energy scale for the coupling between the fluctuator and the CPB and here $\Delta x = x_b - x_a = x_b$ since we have taken $x_a = 0$.

The Hamiltonian given by Eq. (33) can be readily extended to include more charge states of the CPB so that more accurate levels can be found over a larger range of $n_g$. For example, Fig. 2 shows CPB-TLF energy levels as a function of $n_g$ found by computing the energy eigenvalues numerically using three levels of the CPB (n=-1, 0,1). In Fig. 2(a) we set $E_c/h$= 12.48 GHz, $E_b/h$ = 27.04 GHz, $E_r/h$ = 8.32 GHz, $T_{ab}$= 0, and $E_j$= 0, *i.e.* the off-diagonal terms are zero. In this case, we see parabolic energy levels, corresponding to the charging energy of states with well-defined charge on the CPB and position of the fluctuator. For example, the levels labeled $|1a>$, $|0a>$, and $|-1a>$ correspond to the fluctuator being in well *a* and the CPB having excess charge of 2*e*, 0, and -2*e* on the island, respectively. These levels are just the familiar charging energy of curves of the CPB in the limit $E_j$=0. The levels labeled $|1b>$, $|0b>$, and $|-1b>$ are the corresponding levels corresponding to the fluctuator being in well *b*. They are just copies of the $|1a>$, $|0a>$, and $|-1a>$ levels that have been displaced vertically by $E_b$ and horizontally by the induced charge from the fluctuator.

Figure 2(b) shows the corresponding plot of the CPB-TLF energy levels as a function of $n_g$ when $T_{ab}$ and $E_j$ are not set to zero. For this plot we used the same values for $E_c$, $E_b$ and $E_r$ as in Fig. 2(a), but set $E_j/h = T_{ab}/h = 6.24$ GHz. Examination of the plot again reveals similarities to the spectrum of an isolated CPB. Thus, the lowest level is what one would expect for the ground state of the box; this level corresponds predominantly to both the box and the fluctuator being in their ground state and is labeled $|g_b g_f>$. With our choice $x_a = 0$, this level differs little from that of an isolated CPB. Similarly, the sections of the curves labeled $|e_b g_f>$ are very similar to what one would expect for the excited state of the box; this level corresponds predominantly to the box being in its first excited state and the fluctuator being in its ground state. Note in particular the well-known CPB avoided crossing of size $E_j$ between $|g_b g_f>$ and $|e_b g_f>$ at $n_g = \pm 1$.

The excited level labeled $|g_b e_f>$ in Fig. 2(b) requires some additional discussion. The state $|g_b e_f\rangle$ that is responsible for this level has a large amplitude for the box to be in its ground state and the fluctuator to be in its excited state. We note that this curve looks similar to the curve for the level for state $|g_b g_f\rangle$, except that it has been shifted upward along the energy axis and to the right along the $n_g$ axis. The upward shift of the $|g_b e_f\rangle$ curve is approximately just the difference in energy between the ground state and the excited state of the fluctuator excited state, *i.e.* from Eq. (24)-(25) this is about $\sqrt{E_b^2 + 4T_{ab}^2} \approx 30$ GHz, in good agreement with the figure. The shift of the characteristic to the right is caused by charge that is induced on the island when the fluctuator changes from its ground state to its excited state and is just:

$$\Delta n_g = \frac{Q \Delta x}{ed} = \frac{E_r}{2E_c} \qquad [35]$$

*i.e.* the state appears to have an effective gate charge of $n_g$-$\Delta n_g$. For our parameters, this yields $\Delta n_g \approx 0.33$, in good agreement with Figs. 2(a) and 2(b),

A similar situation occurs for the sections of curve labeled $|e_b e_f\rangle$ in Fig 2(b). These sections correspond to both the box being in its first excited state and the fluctuator being in its excited state. The $|e_b e_f\rangle$ sections appear to be a copy of the $|e_b g_f\rangle$ curve that has been shifted upward and to the right, and the amount of these vertical and horizontal shifts is the same as for the shift from the $|g_b g_f\rangle$ to $|g_b e_f\rangle$ sections. Note in particular that the avoided crossing between $|g_b g_f\rangle$ and $|e_b g_f\rangle$ at $n_g = \pm 1$ is replicated between the $|g_b e_f\rangle$ and $|e_b e_f\rangle$ curves at $n_g$ = -0.67 and $n_g$ = 1.33.

Microwave spectroscopy allows direct measurement of the allowed transition spectrum corresponding to differences in the energy levels. Figure 3 shows the calculated spectrum for transitions from the ground state $|g_b g_f\rangle$ to $|e_b g_f\rangle$ (labeled CPB in Fig. 3) and from $|g_b g_f\rangle$ to $|g_b e_f\rangle$ (labeled TLF). Ignoring the small avoided crossings, the section of the curves labeled CPB are essentially just the usual transition spectrum between the ground state and excited state of the CPB. In contrast, the section of the curve labeled TLF in Fig. 3 has a saw-tooth shape which looks quite different from the $|g_b e_f\rangle$ and $|g_b g_f\rangle$ curves in Fig. 2(b) from which it was determined by subtraction. In particular, the TLF curve in Fig. 3 shows a nearly linear variation in transition frequency as a function of $n_g$ between about $n_g = -0.5$ and $n_g = 1$, followed by a rapid reset near $n_g = \pm 1$. The slope of the linear section is just due to the change in the energy of the charge fluctuator in well *b* due to the gate voltage, as expressed by Eq. (17), and the slope scales with $E_r$. The tunneling of the fluctuator also has an impact; large $T_{ab}/E_b$ causes an effective decrease in $Q\Delta x$ and a shallower slope in the TLF spectrum.

The rapid change in the TLF spectrum at $n_g = \pm 1$ is due to a Cooper pair tunneling onto the island, which resets the island potential $V_i$. This reset happens every $\Delta n_g = 2$, and causes the resulting characteristics of the CPB to be periodic along the $n_g$ axis with period 2. Since we assumed the fluctuator is sitting in the ultra-small junction's barrier, between the island and ground, it will be subjected to an electric field that is set by $V_i$, and thus its characteristics will also be periodic in $n_g$ with the same period as the CPB. We note that the characteristics would not be strictly periodic if the fluctuator were sitting in the gate capacitor $C_g$, since in that case the fluctuator would experience a field that was set by $V_g - V_i$, which is not periodic in $n_g$. The implication is that measurements of the spectrum periodicity could thus reveal some information about the location of a fluctuator.

We also note in Fig. 3 the presence of two small avoided crossings in the TLF - CPB spectrum, one near $n_g = -0.3$ and the other near $n_g = 0.5$. The avoided crossing at $n_g = -0.3$ is about 1.2 GHz and is due to coupling between the $|g_b e_f\rangle$ and $|e_b g_f\rangle$ states which for this value of $n_g$ are perturbed from $|0b\rangle$ and $|-1a\rangle$ respectively [see Fig. 2(a)]. The avoided crossing at $n_g = 0.5$ is only about 0.6 GHz and is also due to coupling between the $|g_b e_f\rangle$ and $|e_b g_f\rangle$ states which for this value of $n_g$ are perturbed from $|0b\rangle$ and $|1a\rangle$ respectively. Examination of Fig. 2(b) shows that these avoided crossings are much smaller in size than the well-known CPB avoided crossings at $n_g=\pm1$, which are of size $E_j \sim 8$ GHz. Indeed, the TLF-CPB avoided crossings in Fig. 3 are smaller than both of the off-diagonal terms ($E_j=T_{ab} \sim 6$ GHz) as well as the coupling parameter $E_r \sim 8$ Ghz. This is not surprising because $\langle 0b|H|1a\rangle = \langle 0b|H|-1a\rangle = 0$, as inspection of Eq. (33) reveals so that the coupling between these levels vanishes to first order. In fact, the splitting between the states $|g_b e_f\rangle$ and $|e_b g_f\rangle$ only arises when the effect of the off-diagonal terms in $H$ (*i.e.* $E_j$ and $T_{ab}$) is evaluated to second order in degenerate perturbation theory, thus explaining the small size of these avoided crossings.

The different sizes of the CPB-TLF avoided crossings in Fig. 3 is more complicated to understand. The larger size of the splitting near $n_g=-0.3$ is due to the closeness in energy of the $|0b\rangle$ and $|-1a\rangle$ states to the $|-1b\rangle$ state in this region [see Fig. 2(a)]. In second order perturbation theory, this leads to an amplitude for states $|g_b e_f\rangle$ and $|e_b g_f\rangle$ to be in the $|-1b\rangle$ state, which then couples to $|0b\rangle$ and $|-1a\rangle$ with strengths $T_{ab}$ and $E_j$, respectively, and leads to a significant perturbation and avoided crossing. In contrast, for the avoided crossing between $|g_b e_f\rangle$ and $|e_b g_f\rangle$ near $n_g=0.5$, the $|1b\rangle$ state is relatively far in energy from the $|0b\rangle$ and $|1a\rangle$ states. The resulting levels $|g_b e_f\rangle$ and $|e_b g_f\rangle$ are much closer to being the pure charge states $|0b\rangle$ and $|1a\rangle$, and the resulting avoided crossing has a correspondingly smaller splitting.

**Conclusions**

In summary, we have presented a simple model of charged two-level fluctuators coupled to a Cooper pair box, with the charge fluctuators residing in the dielectric barrier of the CPB's ultra-small junction. We derive a general expression for the Hamiltonian and identify interaction terms that arise between the CPB and each fluctuator and also between a pair of fluctuators. We then considered the specific case of one fluctuator in a CPB and made a further simplifying assumption that the charge fluctuator acted as a two-level system with the charge tunneling between two locations. For this case, we calculated the energy levels and the transition spectrum between the ground state and the two lowest excited states, and identified a distinctive saw-tooth feature in the spectrum that was caused by the fluctuators. We also identified small second-order avoided crossings between the TLF and CPB parts of the spectrum and noted that they are of different sizes.

Finally, we note that since $n_g$ is adjustable experimentally by applying a gate voltage and $E_j$ can be adjusted by applying a magnetic field to the junction, the transition spectrum and splittings predicted by our model can be tested experimentally. In particular, the CPB parameters $E_j$ and $E_c$ can be determined directly from spectroscopy of the CPB transition from $|g_b>$ to $|e_b>$ and the fluctuator parameters $E_b$, $T_{ab}$, and $E_r$ can be determined by fitting the predicted TLF spectrum (see Fig. 3 for example) to the measured spectrum. In fact, these three fluctuator parameters are determined rather directly by three features in the TLF spectrum; the slope of the linear section (which is determined mainly by $E_r$ and $T_{ab}/E_b$), the average energy needed to excite the TLF (which is determined by $E_b$ and $T_{ab}$), and the magnitude of the splittings (which depends on $T_{ab}/E_b$ $E_r/E_c$ as well as the CPB parameter $E_j/E_c$).

In contrast, when a charge fluctuator couples to a Josephson phase qubit the resulting TLF spectrum cannot be measured as a function of gate voltage; the Josephson supercurrent acts to short out static potentials from appearing across the tunnel junction capacitance. Instead, one finds a fixed TLF resonance frequency as a function of the current through the junction. The resulting phase qubit-TLF spectra produce just two features, the energy needed to excite the TLF (which is related to $E_b$ and $T_{ab}$) and the magnitude of the splitting between the TLF and the junction energy levels (which depends on $T_{ab}/E_b$, $E_r/E_c$ as well as $E_j/E_c$). As a result, it is not possible to independently determine $E_r$, $T_{ab}$ and $E_b$ from a phase qubit-TLF spectrum. Similar considerations apply to a flux qubit-TLF spectrum Thus we conclude that measurements of a fluctuator spectrum when it is coupled to a CPB may allow better access to the parameters of the TLF than corresponding measurement of a TLF coupled to a phase qubit.

**Acknowledgements**

We thank C. Sanchez, E. Tiesinga, T. Palomaki, K. Mitra, R. Simmonds, K. Osborn, C. J. Lobb and J. R. Anderson for many helpful discussions of junctions, CPB's, materials, perturbation theory, or fluctuators, and we gratefully acknowledge the generous support of the Joint Quantum Institute, the Center for Nanophysics and Advanced Materials, and the National Security Agency.

# References

[1] P. Dutta, P. Dimon, and P. M. Horn, Phys. Rev. Lett. **43**, 646 (1979).

[2] C. T. Rogers and R. A. Buhrman, Phys. Rev. Lett. 55, 859 (1985).

[3] M. Constantin and C. C. Yu, Phys. Rev. Lett. **99**, 207001 (2007).

[4] Michael Mück, Matthias Korn, C.G.A. Mugford, J.B. Kycia and John Clarke, Appl. Phys. Lett. **86**, 012510 (2005).

[5] F. C. Wellstood, C. Urbina, and John Clarke, Appl. Phys. Lett. **85**, 5296 (2004).

[6] D. J. Van Harlingen, T. L. Robertson, B. L. T. Plourde, P. A. Reichardt, T. A. Crane, and J. Clarke, Phys. Rev. B **70** 064517 (2004).

[7] R. H. Koch, J. Clarke, W. M. Goubau, J. M. Martinis, C. M. Pegrum and D. J. Harlingen, J. Low Temp. Phys. **51**, 207 (1983).

[8] I. Chiorescu, Y. Nakamura, C. J. P. M. Harmans, and J. E. Mooij, Science **299**, 1869 (2003).

[9] Y. A. Pashkin, T. Yamamoto, O. Astafiev, Y. Nakamura, D. V. Averin, and J. S. Tsai, Nature **421**, 823 (2003).

[10] L. J. Geerligs, V. F. Anderegg, J. E. Mooij, Physica B. **165**, 973 (1990).

[11] G. Zimmerli, T. M. Eiles, R. L. Kautz, J. Martinis, Phys. Lett*.*, **61**, 237 (1992).

[12] L. Ji, P. D. Dresselhaus, S. Han. K. Lin, W. Zheng and J. E. Lukens, J. Vac. Sci. Technol. B. **12**, 3619 (1994).

[13] M. Kenyon, C. J. Lobb, and F. C. Wellstood, Jour. Appl. Phys. **88**, 6536 (2000).

[14] M. J. Ferrari, F. C. Wellstood, J. J. Kingston, and J. Clarke, Phys. Rev. Lett. **67**, 1346 (1991).

[15] M. J. Ferrari, M. Johnson, F. C. Wellstood, J. Clarke, D. Mitzi, P. A. Rosenthal, C. B. Eom, T. H. Geballe, A. Kapitulnik, and M. R. Beasley, Phys. Rev. Lett. **64**, 72 (1990).

[16] M. Johnson, M. J. Ferrari, F. C. Wellstood, J. Clarke, M. R. Beasley, A. Inam, X. D. Wu, and T. Venkatesan, Phys. Rev. B **42**, Rapid Comm., 10792 (1990).

[17] F. C. Wellstood, C. Urbina, and J. Clarke, Appl. Phys. Lett. **50**, 772 (1987).

[18] R. H. Koch, D. P. DiVincenzo and J. Clarke, Phys. Rev. Lett. **98**, 267003 (2007).

[19] S. Machlup, J. Appl. Phys. **25**, 341 (1954).

[20] R. W. Simmonds, K. M. Lang, D. A. Hite, S. Nam, D. P. Pappas, and J. M. Martinis, Phys. Rev. Lett. **93**, 077003 (2004)

[21] J. M. Martinis, K. B. Cooper, R. McDermott, M. Steffen, M. Ansmann, K. D. Osborn, K. Cicak, S. Oh, D. P. Pappas, R. W. Simmonds, and C. C. Yu, Phys. Rev. Lett. **95**, 210503 (2005).

[22] L. Tian and R. W. Simmonds, Phys. Rev. Lett. **99**, 137002 (2007).

[23] B. L. T. Plourde, T. L. Robertson, P. A. Reichardt, T. Hime, S. Linzen, C.-E. Wu, and J. Clarke, Phys. Rev. B **72**, 060506(R) (2005).

[24] Z. Kim and B. Palmer, private communication (2007).

[25] V. Bouchiat, D. Vion, P. Joyez, D. Esteve, M. H. Devoret, Phys. Scr. **T76**, 165 (1998).

[26] Y. Makhlin, G. Schön, and A. Shnirman, Rev. Mod. Phys. **73**, 357 (2001).

[27] Y. Nakamura, Yu. A. Pashkin and J. S. Tsai, Nature **398**, 786 (1999).

[28] A. Blais, R. S. Huang, A. Wallraff, S. M. Girvin, R. J. Schoelkopf, Phys. Rev. A **69**, 062320 (2004).

[29] D. Vion, A. Aassime, A. Cottet, P. Joyez, H. Pothier, C. Urbina, D. Esteve, and M. H. Devoret, Science **296**, 886 (2002).

[30] For a discussion of charging energy in small junction devices see for example: *Introduction to Superconductivity*, Second Edition, M. Tinkham, McGraw-Hill, New York, (1996).

[31] D. Song, A. Amar, C. J. Lobb and F. C. Wellstood, IEEE Trans. on Appl. Supercond. **5**, 3085 (1995).

[32] M. Kenyon, J. L. Cobb, A. Amar, D. Song, N. M. Zimmerman, C. J. Lobb, and F. C. Wellstood, J. Low Temp. Phys. **123**, 103 (2001).

**Figure Captions**

Fig. 1. (a) Circuit schematic of Cooper Pair Box (CPB). $C_j$ is the ultra-small tunnel junction capacitance and $C_g$ is the gate capacitance. (b) Charge fluctuators $Q_1$ and $Q_2$ in the dielectric tunnel barrier of the CPB ultra-small junction $C_j$. (c) Charge fluctuator $Q_1$ can tunnel between two locations $x_a$ and $x_b$ in an asymmetric two-well potential.

Fig. 2. Energy levels of a CPB coupled to a fluctuator. In each case a 3-level approximation was used for the CPB ($n$= -1, 0, 1) and the CPB has $E_c/h$= 12.48 GHz and the fluctuator has $E_b/h = 27.04$ GHz and $E_r/h = 8.32$ GHz. (a) Energy levels obtained by setting both the Josephson energy $E_j$ and the fluctuator tunneling term $T_{ab}$ to zero. The resulting levels are the pure position-charge states $|1a>$, $|0a>$, $|-1a>$, $|1b>$, $|0b>$, and $|-1b>$ and there are no avoided level crossings. (b) Solid curves show energy levels found by setting $E_j/h = T_{ab}/h = 6.24$ GHz, with the rest of the parameters the same as in (a). Black curve is the $|g_b g_f>$ level, green is $|e_b g_f>$, red is $|g_b e_f>$, and yellow is $|e_b e_f>$. For comparison, dashed curves in (b) are the same as curves in (a).

Fig. 3. Plot shows calculated transition frequency between the ground state and the two lowest excited states of the coupled CPB - fluctuator system. The CPB and TLF parameters were the same as in Fig. 2(b). The sections labeled CPB (green curves) correspond to excitation of the CPB from its ground state to first excited state, with the fluctuator in its ground state. The sections labeled TLF (red curves) correspond to excitation of the fluctuator from its ground state to first excited state, with the CPB in its ground state. The avoided crossing between the TLF spectrum and the CPB spectrum at $n_g = -0.3$ is about twice as large as the avoided crossing at $n_g= 0.5$.

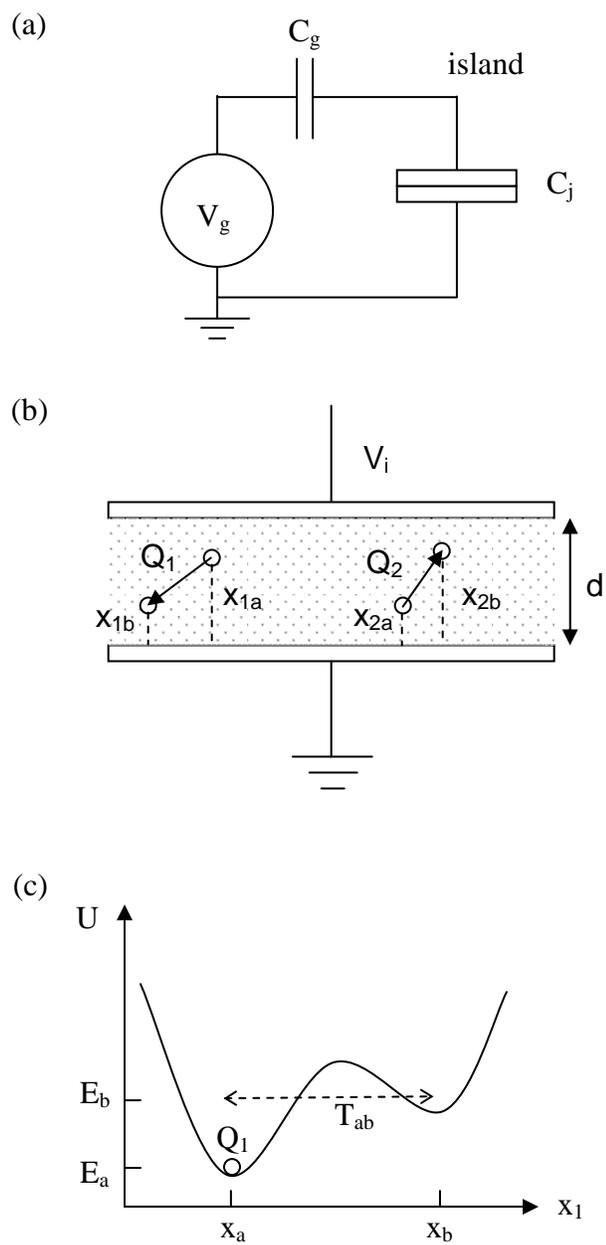

Figure 1. Wellstood *et al.*

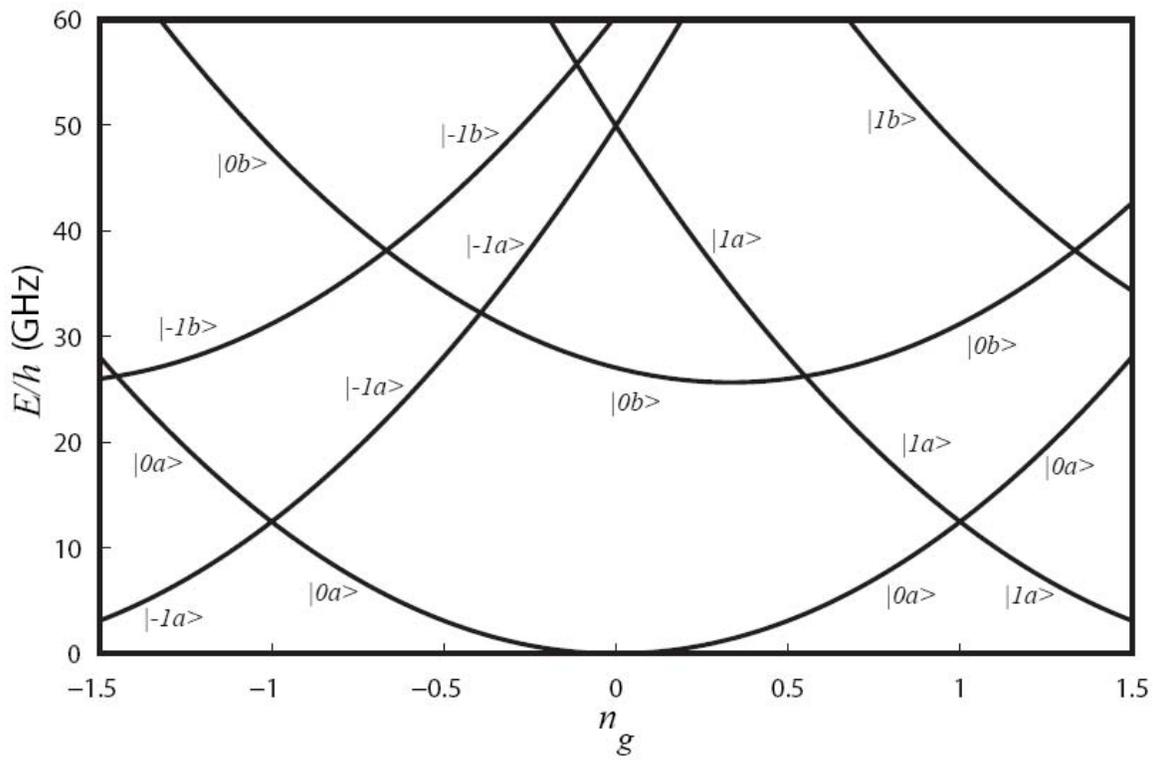

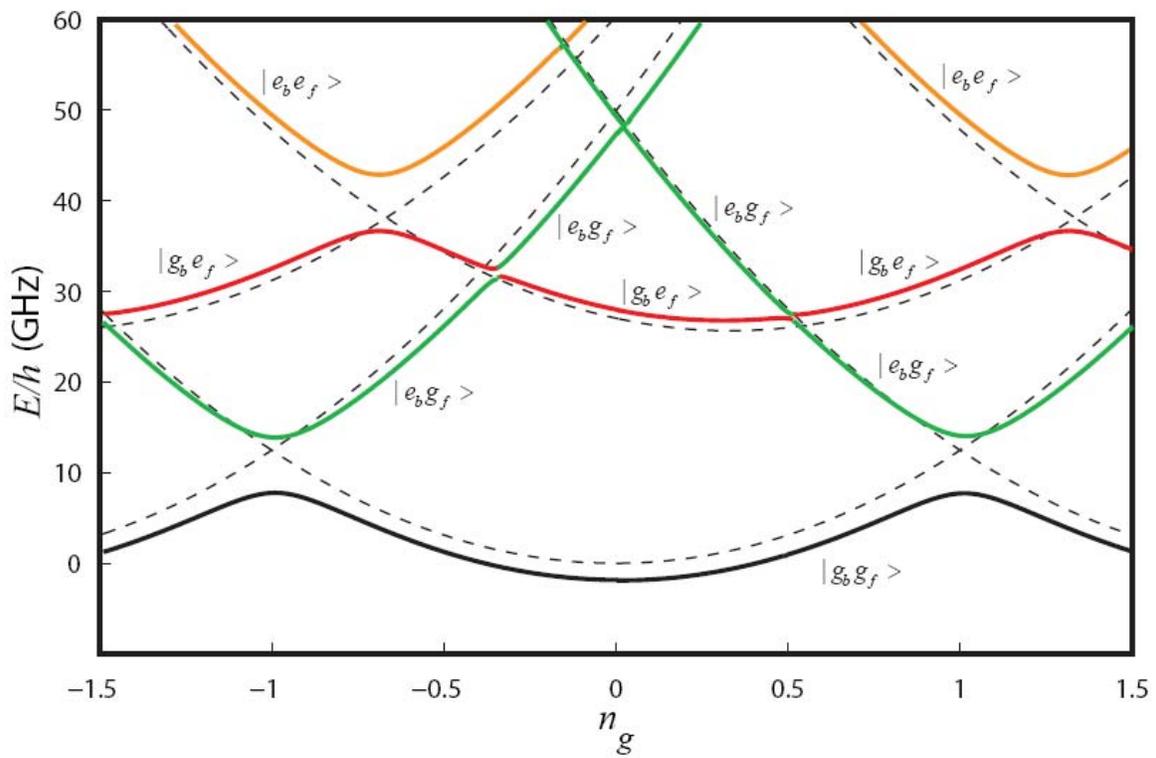

Fig. 2. Wellstood *et al.*

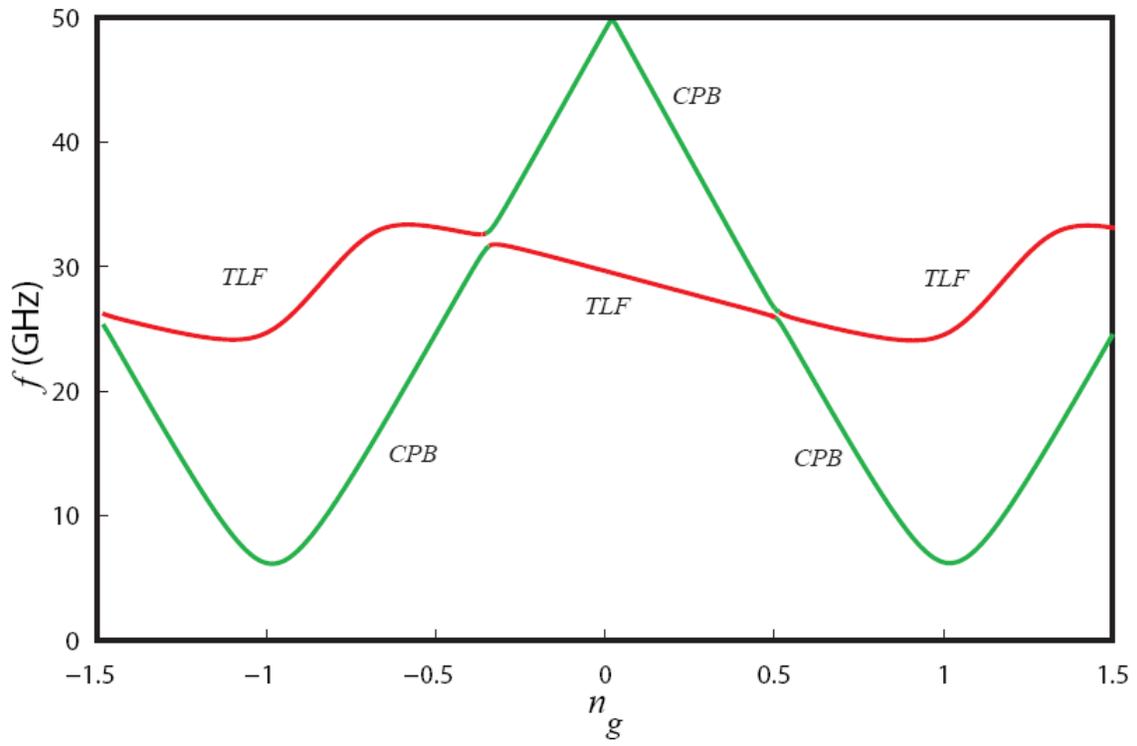

Fig. 3. Wellstood *et al.*